\documentclass[aps,prl,amsmath,amssymb,superscriptaddress]{revtex4}
\usepackage{graphicx}
\usepackage{dcolumn}
\usepackage{color}
\usepackage{bm}
\usepackage{amsmath,amssymb,amsfonts,bbold}
\usepackage[normalem]{ulem}

\newcommand{\beq}{\begin{equation}}
\newcommand{\eeq}{\end{equation}}
\newcommand{\beqa}{\begin{eqnarray}}
\newcommand{\eeqa}{\end{eqnarray}}
\newcommand{\sign}{\text{sign}}

\newcommand{\sgn}{\mathop{\mathrm{sgn}}}
\begin{document}

\title{Mapping the effect of defect-induced strain disorder on the Dirac states  of topological insulators }

\author{Oliver Storz} 
	\address{Physikalisches Institut, Experimentelle Physik II, 
	Universit\"{a}t W\"{u}rzburg, Am Hubland, 97074 W\"{u}rzburg, Germany}	
\author{Alberto Cortijo} 
	\address{Instituto de Ciencia de Materiales de Madrid, CSIC, Sor Juana In\'es de la Cruz 3, Cantoblanco, E-28049 Madrid, Spain}
\author{Stefan Wilfert} 
	\address{Physikalisches Institut, Experimentelle Physik II, 
	Universit\"{a}t W\"{u}rzburg, Am Hubland, 97074 W\"{u}rzburg, Germany}	
\author{K.\,A.\,Kokh} 
       \address{V.S. Sobolev Institute of Geology and Mineralogy, Siberian Branch, Russian Academy of Sciences,
630090 Novosibirsk,  Russia}
\address{Novosibirsk State University, 630090 Novosibirsk,  Russia}
       \author{O.\,E.\,Tereshchenko} 
        \address{Novosibirsk State University, 630090 Novosibirsk,  Russia}
\address{A.V. Rzanov Institute of Semiconductor Physics, Siberian Branch, Russian Academy of Sciences, 630090 Novosibirsk,  Russia}      
\author{Mar\'ia A. H. Vozmediano} 
        \email[corresponding author: ]{vozmediano@icmm.csic.es}
	\address{Instituto de Ciencia de Materiales de Madrid, CSIC, Sor Juana In\'es de la Cruz 3, Cantoblanco, E-28049 Madrid, Spain}        
\author{Matthias Bode} 
	\address{Physikalisches Institut, Experimentelle Physik II, 
	Universit\"{a}t W\"{u}rzburg, Am Hubland, 97074 W\"{u}rzburg, Germany}	
	\address{Wilhelm Conrad R{\"o}ntgen-Center for Complex Material Systems (RCCM), 
	Universit\"{a}t W\"{u}rzburg, Am Hubland, 97074 W\"{u}rzburg, Germany}
\author{Francisco Guinea} 
	\address{IMDEA Nanoscience Institute, E-28049 Madrid, Spain} 
	\address{School of Physics and Astronomy, University of Manchester, Oxford Road, Manchester M13 9PL, UK}    
\author{Paolo Sessi} 
\email[corresponding author: ]{sessi@physik.uni-wuerzburg.de}
	\address{Physikalisches Institut, Experimentelle Physik II, 
	Universit\"{a}t W\"{u}rzburg, Am Hubland, 97074 W\"{u}rzburg, Germany}



\date{\today}

\vspace{1cm}
\begin{abstract}
\vspace{1cm}
\bf{We provide a detailed microscopic characterization of the influence of defects-induced disorder onto the Dirac spectrum of three dimensional topological insulators. By spatially resolved Landau-levels spectroscopy measurements, we reveal the existence of nanoscale fluctuations of both the Dirac point energy as well as of the Dirac-fermions velocity which is found to spatially change in opposite direction for electrons and holes, respectively. These results evidence a scenario which goes beyond the existing picture based on chemical potential fluctuations. The findings are consistently explained by considering the microscopic effects of local stain introduced by defects, which our model calculations show to effectively couple to topological states, reshaping their Dirac-like dispersion over a large energy range.  In particular,  our results indicate  that the presence of microscopic spatially varying stain,  inevitably present in crystals because of the random distribution of defects, effectively couple to topological states  and should be carefully considered for correctly describing the effects of disorder.  }
\vspace{1cm}
\noindent
\end{abstract}


\maketitle
\newpage
The recent discovery of the new class of materials named topological insulators (TIs) 
represents a milestone in condensed matter physics. 
TIs are materials insulating in the bulk but conductive on their surface, 
where they host linearly dispersing gapless states which, 
unlike conventional surface states found in metals and semiconductors, 
cannot be destroyed as long as time-reversal symmetry remains preserved \cite{HK2010,QZ2011}. 
The strong spin-orbit coupling perpendicularly locks the spin to the momentum, 
resulting in a chiral spin-texture which restricts scattering channels \cite{RSP2009,ZCC2009}
and gives rise to spin currents intrinsically tied to charge currents \cite{KVB2007}. 
Beyond their fundamental interest, these unique properties make TIs 
a promising platform for spintronics, magneto-electric, and quantum computing applications \cite{M2010}. 

Immediately after their discovery, it become evident that the presence of defects crucially influences the study of the transport properties of TIs. Defects introduce bulk carriers, thereby complicating the detection of surface conduction channels. Although this problem can be minimized by using thin films instead of bulk crystals, their random distribution leads to the formation of substantial surface disorder, whose effects are far from being completely understood. These are usually described within the framework of the potential disorder created by the local gating effect of defects, which gives rise to charge puddles and spatial fluctuation of the chemical potential making the spin-momentum locking ill-defined over length scales of few nanometers.

Here, we demonstrate the existence of another interaction mechanism between defects and TIs surface states which goes beyond this simple picture. By spatially resolved Landau level spectroscopy measurements, we visualize the existence of substantial spatial fluctuations not only of the Dirac point, as expected on the base of a pure potential disorder picture, but also of the Dirac fermion velocity. Even more remarkably, this is found to change in opposite direction for electrons and holes. Our results are explained by considering the effect of the strain disorder introduced by bulk defects, which effectively couples to topological states reshaping their energy spectrum over a large energy range. The strain picture allows to directly correlate the changes of the Dirac fermion velocity with the fluctuations of the Dirac point, a unique feature of the model.

\begin{figure}[b]   
\includegraphics[width=.95\columnwidth]{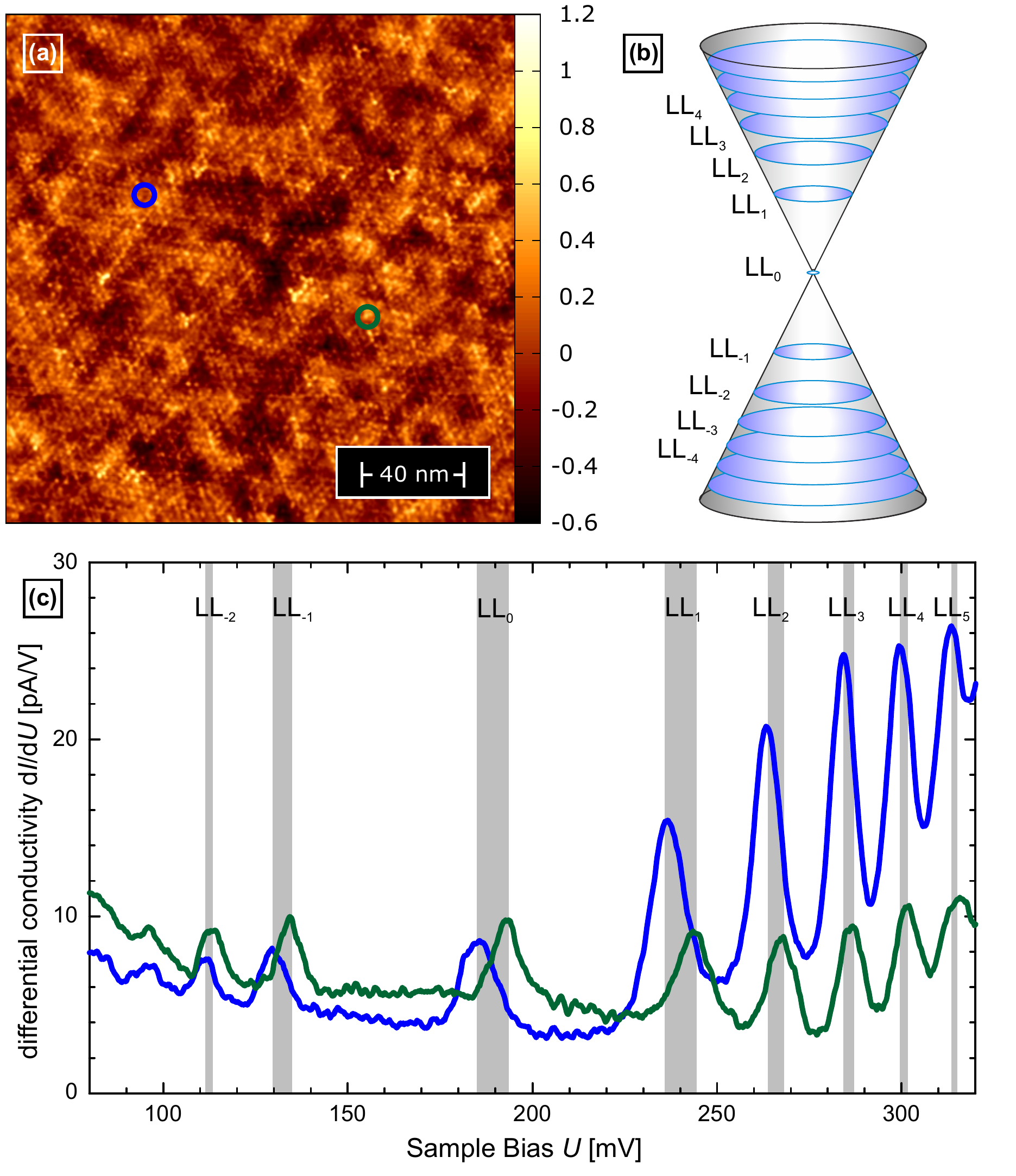}%
\caption{(a) Scanning tunnelling microscopy image acquired in the constant-current mode on Sb$_2$Te$_3$. Several defects introduced during the growth process are visible onto the surface. (b) Schematic illustration of the Landau levels condensation of topological states under high magnetic fields. (c) Scanning tunnelling spectroscopy data acquired over two different sample regions. Green and blue lines correspond to the positions marked by circles in (a).}
\label{Fig:LL}
\end{figure}  

Figure\,\ref{Fig:LL}(a) shows a scanning tunneling microscopy (STM) image 
of the surface of a prototypical binary TI, i.e. Sb$_2$Te$_3$ \cite{SSB2016}. As it will become clear in following, the choice of this material is motivated by the fact that, contrary to Bi$_2$Te$_3$ and Bi$_2$Se$_3$ , Sb$_2$Te$_3$ gives access to both electron- and hole-like dynamics.
Several intrinsic defects 
incorporated during the growth process are visible on the surface \cite{BAS2016}. 
The appearance of these defects strongly depends on the scanning energy. 
As discussed in Ref. \cite{JSM2012} their character, i.e.\ vacancies or antisites, 
as well their exact location within the first Sb$_2$Te$_3$ quintuple layer can be understood 
by carefully analyzing their spatial extension and symmetry.

This surface hosts a 2D electron gas which, based on Ref. \cite{Zetal09}, can be effectively described by the following Dirac Hamiltonian:

\begin{equation}
\mathcal{H}_{\rm eff}= v_{\rm F}\hat{\sigma}_i\cdot \hat{k}_i+D\hat{\sigma}_0 \hat{k}^2,
\label{Heff}
\end{equation}
which is the massless Dirac equation in two dimensions corrected with a band--bending term \cite{ZHetal10}. $\hat{\sigma}_i$ are a set of Pauli matrices, $k$ is the momentum over the surface, and the index $i=1,2$. The Hamiltonian is parametrized by the Fermi velocity $v_{\rm F}$ and the band bending parameter $D$ which accounts for deviations of the band from linearity and plays a crucial role in explaining the electron-hole asymmetry found in the measurements.

The application of strong magnetic fields perpendicular to the surface
leads to the condensation of the 2D topological states into a well-defined sequence 
of Landau levels (LLs) as schematically illustrated in Fig.\,\ref{Fig:LL}(b). Ignoring the band bending term ($D=0$) their energy spacing is given by:
\beq
	E_{n} - E_{\rm D} = \sgn(n) \hbar v_{\rm F} \sqrt{2 e |n| B}, \qquad n = 0, \pm 1, ... ~,
\label{Eq:LL}
\eeq 
from here on we adopt the natural unit  $\hbar =1$, $n$ is the LL index, $B = \mu_0 H$ the magnetic flux density. $E_D$ marks the deviation of the Fermi energy  from the position of the Dirac cone (zeroth LL) due to extrinsic doping, charge impurities, or other effects \cite{li09}. It will also play an important role in our analysis.

As we see the LL distribution provides direct access to the two parameters determining the electrons dynamics of TIs:
(i) the position of the Dirac point $E_D$ and (ii) the Fermi velocity $v_{\rm F} $\cite{CSZ2010,JWC2012,OSL2013} which can all be experimentally obtained by  STM spectroscopic measurements. We would like to note that,  
although these properties are in principle also detectable by  other techniques, 
such as angular resolved photoemission (ARPES),  
the use of local probes allows to visualize the existence and the impact of local phenomena 
which will would remain undetected using spatially averaging techniques.

Figure\,\ref{Fig:LL}(c) reports scanning tunneling spectroscopy (STS) measurements 
performed at a magnetic field of $B = 12$\,T.
The emergence of well-defined peaks 
signals the condensation of the topological states into Landau levels. 
The green and the blue spectra were acquired by positioning the tip 
over two different sample positions marked by green and blue circles in Fig.\,\ref{Fig:LL}(a), respectively. 
Their direct comparison reveals significantly different LL positions in the two sample areas. Even more interestingly, 
they clearly reveal the existence of different energy shifts for each one of the Landau level,
as highligthed by the shadowed gray boxes in Fig.\,\ref{Fig:LL}(c).
These spectroscopic variations go beyond a simple picture based on defects-induced nanoscale spatial fluctuations of the chemical potential\cite{BRS2011}.  Indeed, if this were the case, spectra acquired in different positions would be simply rigidly shifted with respect to each other, an expectation in obvious disagreement with our experimental results,

\begin{figure*}[b]   
\includegraphics[width=.9\textwidth]{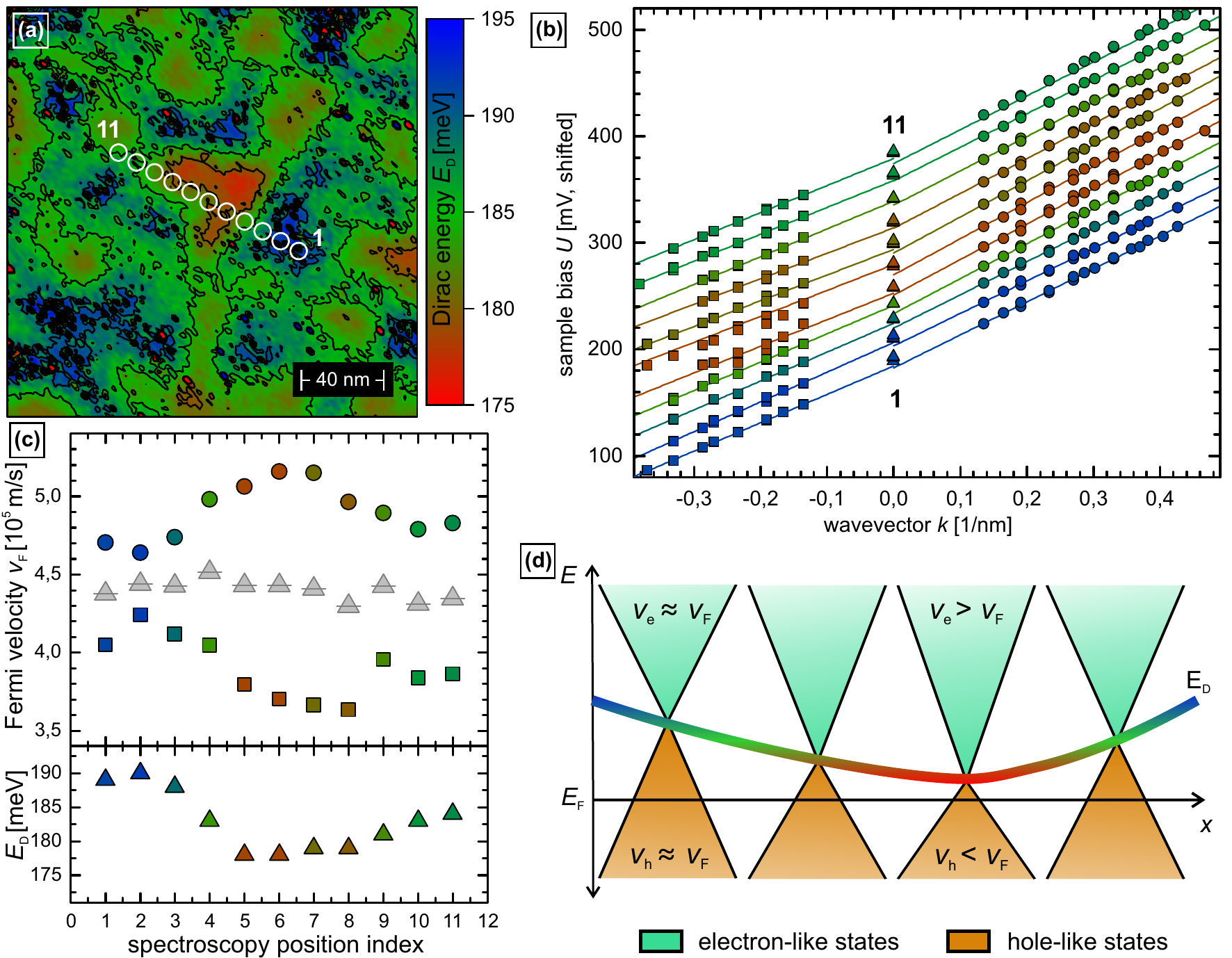}%
\caption{(a) Dirac point mapping with respect to the Fermi level obtained over the sample region displayed in Fig. 1(a). (b) Landau levels positions for points taken along the trace identified by circles in (a). The results obtained over each point have been vertically shifted to improve clarity; (c) Dirac fermions velocity for electrons (circle) and holes (square) branches of the Dirac cone obtained by fitting the data presented in (b). The gray triangles report the average velocity, which stays constants for all positions. The bottom panel quantifes the position of the Dirac point with respect to the Fermi level. A comparison between upper and lower panels reveals that the electron-hole asymmetry increases once the Dirac point moves closer to the Fermi level; (d) schematic illustration of the Dirac point fluctuations and electron-hole symmetry breaking.  }
\label{Fig:Renorm}
\end{figure*}  

To systematically investigate this trend, full spectroscopic measurements 
have been performed over the entire sample region shown in Fig.\,\ref{Fig:LL}(a). 
By acquiring STS curves for each point in the image, the existence of local variation 
of the Dirac fermions can be visualized down to the single-atomic level. 
Since the zeroth LL is locked to the Dirac point, spatially mapping its position 
allows to visualize its fluctuations 
as illustrated in Fig.\,\ref{Fig:Renorm}(a).  
Although always located above the Fermi level due to the intrinsic p-doping of Sb$_2$Te$_3$ \cite{JSM2012}, 
the Dirac energy $E_{\rm D}$ is found to spatially fluctuate by of approximately 20\,meV. 

To visualize how higher order LL peaks are influenced by these fluctuations, STS measurements have been performed along the trace identified by circles in Fig.\,\ref{Fig:Renorm}(a). Note that this trace extends over extreme values of $E_{\rm D}$, thereby spanning all possible values of the Dirac point with respect to $E_{\rm F}$.  
The exact energy positions of LLs have been obtained by considering the maximum of a Gaussian function fitting each peak.
Results are reported in Fig.\,\ref{Fig:Renorm}(b) revealing that, while electron- and hole-like branches of the Dirac cone have similar slopes in regions where the Dirac point is well above the Fermi (blue lines), a kink signaling the breaking of the electron-hole symmetry becomes evident once the Dirac point is closer to the Fermi (red line).  Note that electron--hole asymmetries have been previously reported in the Dirac spectrum of TIs \cite{JWC2012,pauly15}. Their existence is not surprising and can be traced back to density functional theory data \cite{PBL2012}.  However, our spatially resolved measurements reveal that: (i) this asymmetry is not constant but fluctuates over the nanoscale, (ii) it is directly correlated to the fluctuations of the Dirac point.  

A linear fit of the data allows to quantify the strength of this effect in terms of the changes induced onto the Dirac fermions velocity. Results are summarized in Fig.2(c). While the average fermions velocity stays locked at a value of approximately $v_{F}=4.3\cdot 10^{5}m/s$, in agreement with earlier reports \cite{JWC2012,PBL2012}, electrons and holes velocities are found to be symmetrically enhanced and reduced by up to 15\% of their original value.  A comparison of upper and lower panels in (c) allows to directly and unequivocally link the observed behavior to the position of the Dirac energy $E_{\rm D}$ with respect to the Fermi level $E_{\rm F}$, with the effect being stronger the smaller $E_{\rm D}- E_{\rm F}$ \cite{Note}.  
These results are schematically summarized in Fig.\,\ref{Fig:Renorm}(d).  

Tip-induced band-bending effects, reported in Refs. \cite{CSZ2010,JWC2012} cannot account for the observed behavior. Indeed, if this were the case,  the deviation from a linear trend would monotonically depend from the tip-sample  bias, i.e. on the electric field applied within the tunneling junction. Obviously this picture is not consistent with our observations.  Furthermore, because of the well know $p$--type doping of Sb$_2$Te$_3$, the electron- and hole-like parts of the Dirac spectrum appear both at positive voltages, such that the electric field has always the same polarity. Consequently, tip-induced band-bending effects cannot explain the existence of spatial fluctuations of the Dirac velocity of opposite sign for electrons and holes.
To further exclude this effect, STS measurements have been performed by significantly changing the stabilization current. This results in a variation of the tip-sample distance, thereby changing the electric fields across the junction. As shown in the supplementary materials, the position of the LLs is not affected by the different set-points. 
We can thus safely conclude that our observation are an intrinsic property of the system.

We are then faced to explain local changes on the Fermi velocity, on the position of the zeroth LL (which marks the deviation of the Fermi energy from the position of the Dirac point), and a pronounced electron--hole asymmetry.
Similar local variations have been previously detected in graphene \cite{li09,Stroscio2} and attributed to charge inhomogeneities induced by the substrate or by charge impurities  and to the Fermi velocity renormalization due to Coulomb interactions \cite{GGV94,EGetal11}.  These effects, which affect the  Fermi velocity and shift the Dirac point in an uncorrelated way, can not account for the dependence found in our data. Moreover, the effect of Coulomb interactions is expected to be much smaller in topological insulators than in graphene since  
the dielectric constant is bigger. The logarithmic renormalization of the Fermi velocity due to Coulomb interactions   does not fit the magnitude nor the trend of our data leaving strain as the most plausible explanation.

We suggest that the strain created by bulk lattice deformations due to defects, misalignment, and stacking faults can effectively account for all the features observed at the surface of Sb$_2$Te$_3$. In what follows we sketch the main lines of this proposal. A complete derivation of the effect of  strain on the low energy effective Hamiltonian  at the surface of a topological insulator is provided in the  Supplementary Material.


Simple symmetry arguments \cite{MJSV13} allow to construct the most general effective Hamiltonian coupling the lattice deformations in the bulk parametrized by the strain tensor $u_{ij}=\frac{1}{2}(\partial_i u_j+\partial_j u_i);\; i,j=1,2,3$, to the electronic degrees of freedom. For the material's lattice and after projecting to the surface, the Hamiltonian in \eqref{Heff} is modified by the strain dependent  terms ($Tr(\bar{u})$ is the trace of the strain tensor $\bar{u}_{ij}$):
 \beq
 \mathcal{H}[u]_{surf}=\lambda_{V}\hat{\sigma}_{i}\bar{u}_{ij}(\bm{r})\hat{k}_{j}+\lambda_{V'}Tr(\bar{u})(\bm{r})\hat{\sigma}_{i}\hat{k}_{i}+
 \lambda_{0}Tr(\bar{u})(\bm{r})\sigma_{0}+\lambda_{D_{1}}Tr(\bar{u})(\bm{r})\delta_{ij}\hat{k}_i\hat{k}_j\sigma_{0}+\lambda_{D_{2}}\bar{u}_{ij}(\bm{r})\hat{k}_{i}\hat{k}_{j}\sigma_{0}
 .\label{HUeff}
 \eeq
The most important magnitude is  the effective two dimensional strain tensor defined as \begin{equation}
 \bar{u}_{ij}(\bm{r})=\int^{\infty}_{-\infty} u_{ij}(\mathbf{r},z)|f(z)|^{2} dz,
 \end{equation}
where $f(z)$ is the envelope function of the surface state \cite{QZ2011}.

The  parameters $\lambda_i$ are
related to the elastic properties of the bulk and remain arbitrary in the symmetry approach although they could be determined within a given microscopic model (tight binding or alike).  The largest in magnitude will be the parameter $\lambda_0$ associated to the deformation potential, a term that induces local variations of the electronic density proportional to the trace of the strain tensor. The remaining parameters
will be  reduced to two in the data analysis: one provides local corrections to the Fermi velocity and the other one induces local corrections to the band bending $D$. 
As we will see, the analysis of our data allows us to assign a value to these parameters which are hard to determine experimentally. 
It is worth to mention that, when comparing with the well known case of graphene,  the so-called elastic gauge fields are absent because our Dirac points lie at the $\Gamma$ point of the Brillouin zone. 

\begin{figure*}[h]
\includegraphics[width=\textwidth]{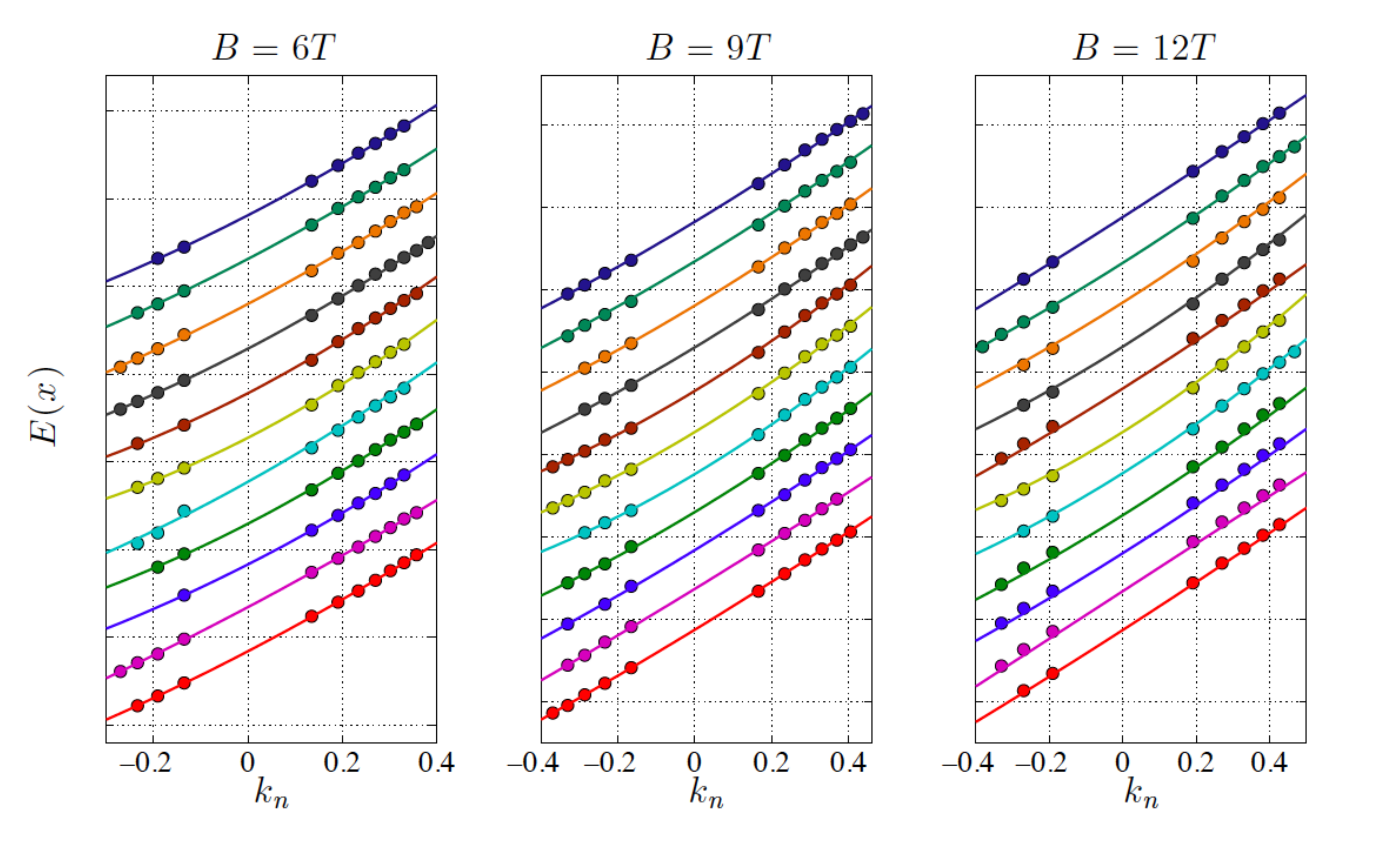}
\caption{Landau level energies as a function of the momenta $k_n$ defined in the text,
measured at three different magnetic fields at the eleven spatial positions depicted in Fig. \ref{Fig:Renorm}. The energies have been shifted upwards for clarity and the position of the zeroth LL has been removed.}
\label{Fig3}
\end{figure*}

Including the magnetic field amounts to substitute $\hat{k}_i$ by the generalized momenta $\pi_{i}=i\partial_{i}-eA_{i}$ in \eqref{HUeff}.
In perturbation theory 
we get the Landau level energies:
\beqa
E_{0}=\bar{E}+\frac{1}{2}Dl^{-2}_{B}+\lambda_{0}Tr(\bar{u})+\frac{1}{2}\bar{\lambda}_{D}l^{-2}_{B}Tr(\bar{u}),\label{zLLenergy}
\eeqa
\beq
E_{n}=\bar{E}+\lambda_{0}Tr(\bar{u})+k_{n}(v_{F}+\bar{\lambda}_{V}Tr(\bar u))+
k^{2}_{n}(D+\bar{\lambda}_{D}Tr(\bar u)).\label{LLenergies}
\eeq
where $k_{n}=\sign(n)\sqrt{|n|}l^{-1}_{B}$,  $l_B$ is the magnetic length, 
and $\bar{\lambda}_{V}=\lambda_{V_{1}}+\frac{1}{2}\lambda_{V_{2}}$, $\bar{\lambda}_{D}=\lambda_{D_{1}}+\frac{1}{2}\lambda_{D_{2}}$.
As we anticipated, the LL energies depend only on three elastic parameters.
As we can see, all the strain dependence of the perturbative results for the Landau level energies  is through the trace of the strain tensor $Tr(\bar{u}(\bm{r}))$. This  allows us to correlate the Landau level energies with the energy of the zeroth Landau level at different points, giving a space-independent check of the theory. 
The Landau level energies were measured at eleven different positions $\mathbf{r}_{i}$ of the sample and at three different magnetic fields, $B=6, 9$ and 12  T. These energies are represented in Fig. \ref{Fig3} as a function of the effective momentum $k_n$. The energies of the same Landau level at different points have been shifted artificially upwards for clarity. 
%

From the fitting of these energies we can extract the parameters
\beq
\bar{v}_{F}(\mathbf{r}_{i}) =v_{F}\left(1+\frac{\bar{\lambda}_{V}}{v_{F}}Tr(\bar{u}(\mathbf{r_{i}}))\right),\quad
\bar{D}(\mathbf{r}_{i})=D\left(1+\frac{\bar{\lambda}_{D}}{D}Tr(\bar{u}(\mathbf{r}_{i}))\right),
\eeq
for the eleven positions and the three different values of the magnetic field.

\begin{figure*}[h]
\includegraphics[width=\textwidth]{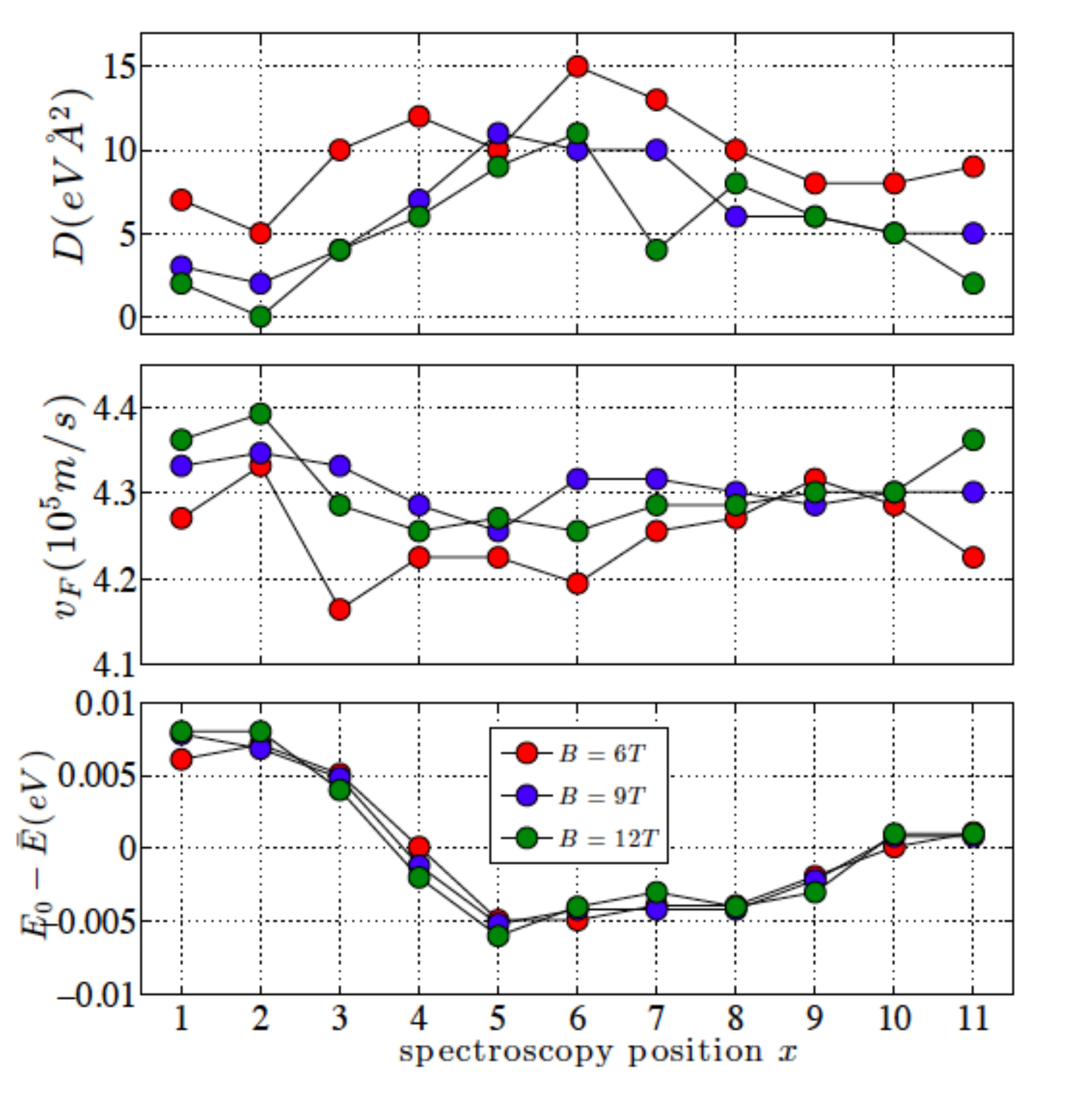}
\caption{ Estimated values of the parameters $\bar D$ (upper panel) and $\bar v_{F}$ (middle) for the various positions in the sample shown in Fig. \ref{Fig:Renorm} (a). The values are extracted from  eqs. \eqref{par}  and shown for  magnetic fields B=6, 9, 12 T. Despite the dispersion in the values of the parameter $D$, the coincidence of the trend for the different values of the magnetic field constitutes a very non trivial test of the strain model. The bottom panel shows the variation of the Fermi energy at the different positions of the sample for different values of the magnetic field. The coincidence is remarkable.}
\label{Fig4}
\end{figure*}

Assuming that  $\bar{\lambda}_{D}$ is much smaller than $\lambda_0$ (a condition verified a posteriori from the data analysis as can be seen in Table 1 of the Supplementary Material), we can linearly correlate the fitting parameters obtained from the Landau level energies with the energy of the zeroth Landau level what allows us to extract phenomenological values for the electron--phonon coupling in the bulk material:
\beq
\bar{v}_{F,i} =v_{F}+\frac{\bar{\lambda}_{V}}{\lambda_{0}}\left(E_{0}-\bar{E}\right), \qquad
\bar{D}_{i} =D+\frac{\bar{\lambda}_{D}}{\lambda_{0}}\left(E_{0}-\bar{E}\right).\label{par}
\eeq
%
The correlations are plotted in Fig. \ref{Fig4}(a,b). 
The mean Fermi velocity averaged from the three values is 
$v_{F}=4.3\cdot 10^{5}m/s$, in remarkably good agreement with other experimental reports \cite{JWC2012,PBL2012,SRB2014}. The mean value for the band-bending parameter is $D=6.9$ eV\AA$^2$ is of the same order of magnitude of the experimental values reported for a similar compound \cite{ZHetal10}. This is an important consequence of the analysis.


A comparative analysis of the electron-hole asymmetry in graphene generated by strain and charge-defect scattering was done in \cite{BW15}. They see that the Fermi velocity  of the electrons  measured through the LL slope is higher than that of the holes in the strain model while the opposite asymmetry is found in the charged defect explanation. Our results follow the strain behavior. 
The dependence of the parameter D with the position of the sample is another non-trivial test of the strain model that would not occur if the observed anisotropies were simply due to charge inhomogeneities and Coulomb interactions. 

In summary, we detected the existence of subtle local variations of the electron dynamics at the surface of a topological insulator.  Our experiments reveal correlated nanoscale fluctuations of both the Dirac point energy as well as of the Dirac-fermions velocity which, by keeping its average value constant, is found to spatially change in opposite direction for electrons and holes.  These findings are consistently explained within a very simple strain model which effectively describes the local changes in the Dirac fermions velocity and correlates them with the fluctuations of the Dirac point.  As a bonus, our analysis allows to obtain a phenomenological determination of the electron--phonon coupling strength in the bulk material. The linear relation between the energies of higher Landau level and the lowest Landau level at each point is parameter-independent and constitutes a very non--trivial test of the strain model. 
Our study implies that the local strain introduced by defects is a central ingredient which needs to be considered to effectively describe disorder in TIs. 

\section*{Acknowledgments.} 

This research was supported in part by the Spanish MECD grants FIS2014-57432-P, the  European Union structural funds and the Comunidad de Madrid MAD2D-CM Program (S2013/MIT-3007), by the National Science Foundation under Grant No. NSF PHY11-25915,  by the European Union Seventh Framework Programme under grant agreement no. 604391 Graphene Flagship, FPA2012-32828, and ERC, grant 290846. The experimental work was supported by DFG (through SFB 1170 "ToCoTronics"; project A02). MAHV acknowledges useful conversations with Fernando de Juan.



\bibliography{Strain_References}

\begin{thebibliography}{10}
\bibitem{Note}
\bibinfo{title}{Note that, once defects or dopants are present, the separation
  between the local surface dirac point and the fermi $e_{\rm d}- e_{\rm f}$
  can not be considered a good measure of the doping level. indeed, like for
  every semiconductor, bulk dopants and defects shift the position of the fermi
  level with respect to the conduction and valence bands, which are
  well-defined bulk properties. the dirac spectrum and consequently the dirac
  point energy are surface properties which, as described in the manuscript,
  can not be considered as a good "reference frame" but are both significantly
  dressed by defects.} .
\end{thebibliography}

\end{document}